\documentclass[aps,
preprint,
amsmath,amssymb,
groupedaddress
]{revtex4-2}

\usepackage{hyperref}
\usepackage{multirow}
\hypersetup{
    colorlinks=true,
    linkcolor=blue,
    filecolor=blue,      
    urlcolor=blue,
    citecolor=blue}

\begin{document}


\title{Jordan and Einstein frames Hamiltonian analysis\\ for FLRW Brans-Dicke theory}



\author{Matteo Galaverni}
\email[]{matteo.galaverni@gmail.com}
\affiliation{Specola Vaticana (Vatican Observatory), V-00120, Vatican City State}
\affiliation{INAF/OAS Bologna, via Gobetti 101, I-40129 Bologna, Italy\\
ORCiD: 0000-0002-5247-9733}

\author{Gabriele Gionti, S.J.}
\email[Corresponding author: ]{ggionti@specola.va}
\affiliation{Specola Vaticana (Vatican Observatory), V-00120, Vatican City State}
\affiliation{Vatican Observatory Research Group, Steward Observatory, The University Of Arizona,
933 North Cherry Avenue, Tucson, Arizona 85721, USA}
\affiliation{INFN, Laboratori Nazionali di Frascati, Via E. Fermi 40, 00044 Frascati, Italy.\\
ORCiD: 0000-0002-0424-0648}


\date{\today}

\begin{abstract}
We analyze Hamiltonian equivalence between Jordan and Einstein frames considering a mini-superspace model of flat Friedmann-Lema\^{i}tre-Robertson-Walker (FLRW) Universe in Brans-Dicke theory. Hamiltonian equations of motion are derived in the Jordan, Einstein, and in the anti-gravity (or anti-Newtonian) frames. We show that applying the Weyl (conformal) transformations to the equations of motion in the Einstein frame we did not get the equations of motion in the Jordan frame. Vice-versa, we re-obtain the equations of motion in the Jordan frame applying the anti-gravity inverse transformation to the equation of motion in the anti-gravity frame.
\end{abstract}

\maketitle

\section{\label{Jordan-Einstein} Introduction}
The Brans--Dicke theory \cite{Brans1961} is a special case of scalar-tensor theory \cite{Dyer}: 
\begin{eqnarray}
S&=&\int_{M}d^{4}x\sqrt{-g}\left(\phi\;{}^{4}R-\frac{\omega}{\phi}g^{\mu\nu}\partial_{\mu}\phi \partial_{\nu} \phi -U(\phi)\right)\nonumber \\
&+& 2\int_{\partial M} d^3x \sqrt{h}\phi K\;.
\label{BDaction}
\end{eqnarray}

The 
equations of motion for the metric tensor $g_{\mu\nu}$ are: 
\begin{eqnarray}
R_{\mu \nu }-\frac{1}{2}g_{\mu \nu }R&=&\frac{\omega}{\phi^2}\left[\partial_\mu\phi\partial_\nu\phi-\frac{1}{2}g_{\mu\nu}g^{\alpha\beta}\partial_\alpha
 \phi \partial_\beta \phi)\right]+\nonumber\\
 &+&\frac{1}{\phi}\left[\nabla_\mu\nabla_\nu\phi-g_{\mu\nu}\Box \phi -\frac{1}{2}g_{\mu\nu}U(\phi)\right]
 \label{equationforg},
\end{eqnarray}
whereas the equation of motion for $\phi$ is: 
\begin{equation}
(3+2\omega)\Box \phi =\phi \frac{dU}{d\phi}-2U(\phi)\,.
\label{eqstophi}
\end{equation}

In two papers \cite{Gionti2021,galaverni2021jordan}, we have studied in detail the Hamiltonian theory related to the action \eqref{BDaction}, 
{where we focused on the issue of canonical transformations.
Here, we remind that a transformation $(Q^i(q,p), P_i(q,p))$ between two sets of variables $(q^i,p_i)$ and $(Q^i,P_i)$  is canonical if the \textquotedblleft symplectic two form \textquotedblright
$\omega=dq^i \wedge dp_i$ is invariant;
that is, $\omega=dQ^i \wedge dP_i$. This is equivalent to stating that the Poisson brackets fulfill the following conditions:
\begin{eqnarray}
\label{canonicalone}
\{Q^{i}(q,p),P_{j}(q,p)\}_{q,p}&=&\delta^{i}_{j}\,, \label{symplectic}  \\
\{Q^{i}(q,p),Q^{j}(q,p)\}_{q,p}&=&\{P_{i}(q,p),P_{j}(q,p)\}_{q,p}=0\, .
\nonumber
\end{eqnarray}

This implies that the equations of motion are: 
$\dot{Q}^i=\left\{Q^i,H\right\}$ and $\dot{P}_i=\left\{P_i,H\right\}$,
where $H(Q^i,{P}_i)$ is the Hamiltonian function transformed in the new set of canonical variables $(Q^i(q,p), P_i(q,p))$ \cite{arnol2013mathematical}.
}

We have introduced Jordan and Einstein frames \cite{Dicke,Faraoni2006,Dyer} that are defined starting from the Brans--Dicke theory and the Hamiltonian transformations between the two frames, both for the case of $\omega\neq -\frac{3}{2}$ and $\omega= -\frac{3}{2}$ \cite{Olmo}. We have found that these transformations are not Hamiltonian canonical transformations. 
Hamiltonian canonical transformations exist; they are the anti-gravity or anti-Newtonian transformations \cite{Niedermaier2019,Niedermaier2020,Zhang2011,Zhou2012},
but they are not Weyl (conformal) transformations of the metric tensor. Therefore, the solutions of the equations of motion in one frame are not necessarily the solutions of the equations of motion in the other frame. Here, we show this in an even clearer way by considering, in the Brans--Dicke theory, a flat FLRW minisuperspace model. We will study both the cases of $\omega\neq -\frac{3}{2}$ and $\omega= -\frac{3}{2}$. For each case, we derive the Hamiltonian functions and the equations of motion in the Jordan frame and in the Einstein frame.  
We apply the transformations from the Jordan to the Einstein frames on the equations of motion;
that is, we apply the transformations on the equations of motion in the Einstein frame to go back to the Jordan frame. We see that we do not obtain the original equations of motion in the Jordan frame, which means that the transformations from the Jordan frame to the Einstein frame are not canonical transformations.
We notice that only if we employ the anti-gravity (or anti-Newtonian) transformations that the previous procedure reproduces, in the Jordan frame, do we derive the same equations of motion as those from the original Hamiltonian. 

We consider the high symmetrical case of the flat FLRW universe with spatial curvature $k=0$:
\begin{equation}
ds^{2}=-N^{2}(t) dt^{2}+a^{2}(t)dx^3.
\label{FLRWpiatta}
\end{equation}
We derive the Lagrangian function ${\mathcal{L}}$  of this mini-superspace model by substituting this metric in the action \eqref{BDaction} (cfr. \cite{Bonanno2017}):
\begin{equation}
{\mathcal{L}}=-\frac{6a{\dot{a}}^2}{N}\phi-\frac{6a^2{\dot{a}}}{N}{\dot{\phi}}+\frac{\omega a^3}{N\phi}(\dot{\phi})^2-Na^3U(\phi)\,. 
\label{lagra}
\end{equation}

The \textquotedblleft configuration\textquotedblright variables are: the lapse $N=N(t)$, the scale factor of the universe $a(t)$ of the FLRW metric, and the field $\phi=\phi(t)$, which now depends only on time $t$ for symmetry reasons. 

We have organized this essay in four sections. Section \ref{sectII} deals with the Dirac's constraint analysis in the Jordan frame for $\omega\neq -\frac{3}{2}$ and $\omega= -\frac{3}{2}$. 
The anti-gravity (or anti-Newtonian) transformations are introduced in Section \ref{sectIII}. 
The Hamiltonian transformations from the Jordan frame to the Einstein frame are discussed in Section \ref{sectIV}. 
Section \ref{sectV} ends \mbox{with conclusions.  }

\section{Constraint Analysis and Equations of Motion in the Jordan Frame}
\label{sectII}
Given the mini-superspace Lagrangian $\mathcal{L}$ \eqref{lagra}, we can define the ADM Hamiltonian~\cite{ADM} ${\mathcal{H}}_{ADM}$, as has been extensively shown in \cite{Gionti2021,galaverni2021jordan}.
We start with the definition of the canonical momenta \cite{dirac1966,DeWitt1967,Esposito1992}:
\begin{eqnarray}
\pi_N &\equiv&\frac{\partial {\mathcal L}}{\partial \dot{N}}\approx 0\,, \\
\pi_{a}&\equiv&\frac{\partial {\mathcal L}}{\partial \dot{a}} =-\frac{12a{\dot{a}}}{N(t)}\phi(t)-\frac{6a^2}{N(t)}{\dot{\phi}(t)}   \, ,  \\
\pi_\phi&\equiv&\frac{\partial {\mathcal L}}{\partial \dot{\phi}}=-\frac{6a^2{\dot{a}}}{N(t)}+\frac{2\omega a^3}{N\phi(t)}\dot{\phi}(t)\,. 
\label{pippo2}
\end{eqnarray}

The determinant of the Hessian matrix $\frac{\partial^2 {\mathcal{L}}}{\partial \dot{q}^i \partial \dot{q}^j}$ has a zero due to definition of the momentum
$\pi_{N}$ conjugated to $N$. It can have another zero for $\omega=-\frac{3}{2}$, as it is easy to check if we compute the determinant of the sub-Hessian matrix associated to the variable $\dot{a}$ \mbox{and $\dot{\phi}$:}
\begin{equation}
\label{Hess}
\mathrm{det}_{(\dot{a}\dot{\phi})}\left|\frac{\partial^2 {\mathcal{L}}}{\partial \dot{q}^i \partial \dot{q}^j} \right|
= -\frac{12 (2\omega+3)a^4}{N^2}\,.
\end{equation}

Therefore, we discuss the two cases of $\omega\neq-3/2$ and $\omega=-3/2$ separately.

\subsection{$\omega\neq-3/2$ Case}
\label{2.1}

The canonical Hamiltonian is:
\begin{eqnarray}
H_{C}\equiv p_i\dot{q}^i-\mathcal{L} =N\bigg[-\frac{\omega{\pi}^{2}_{a}}{12a\phi (2\omega +3)}-\frac{ {\pi}_{a} \pi_{\phi}}{2 a^2 (2\omega +3)} 
+\frac{ \phi {\pi}^2_{\phi}}{2 a^3 (2\omega +3)} + a^3 U(\phi) \bigg] \ ,
\label{hamiltopunto}
\end{eqnarray}
the Hamiltonian constraint $H$ is the quantity under square brackets.

We introduce an effective Hamiltonian defined as:
\begin{equation}
 H_{\mathrm{E}}\equiv N H + \lambda_N \pi_N   
\end{equation}
where $\lambda_N(q^i, p_i)$ is a Lagrange multiplier.
In order to preserve the primary constraint $\pi_N\approx 0$,
we require:
\begin{equation}
\dot{\pi}_N=\left\{ \pi_N,  H_{\mathrm{E}} \right\}=-H\approx 0
\end{equation}
and obtain a secondary constraint, which is the usual Hamiltonian constraint $H$.

This constraint is preserved since:
\begin{equation}
\dot{H}=\left\{ H,  H_{\mathrm{E}} \right\}\approx 0\,.   
\end{equation}

One can easily see that the two constraints $\pi_N \approx 0$ and $H \approx 0$ are first class, since $\{\pi_N,H\}=0$.

The total Hamiltonian is: 
\begin{equation}
H_{\mathrm{T}}\equiv  H_{\mathrm{E}} + \lambda_H H=  N H + \lambda_N \pi_N + \lambda_H H 
\end{equation}
since $\lambda_H$ and $N$ are two arbitrary multipliers, we can write this in a more simple way:
\begin{equation}
H_{\mathrm{T}}=N H  + \lambda_N \pi_N =H_{\mathrm{E}} 
\end{equation}

We found, in this $\omega\neq-3/2$ case, a primary constraint ($\pi_N\approx 0 $)
and a secondary constraint ($H\approx 0 $);
 they are both first class constraints, since $\left\{\pi_N,H\right\}=0$.

The equations of motion, $\dot{q}^i=\left\{q^i,H_{\mathrm{T}}\right\}$ and $\dot{p}_i=\left\{p_i,H_{\mathrm{T}}\right\}$,  in the Jordan frame for $\omega\neq-3/2$ are:
\begin{eqnarray}
\label{eq_neq32_N}
&& \dot{N}\approx \lambda_N \;,\\
\label{eq_neq32_pi_N}
&&\dot{\pi}_N=-H\approx 0\,, \\
\label{eq_neq32_a}
&& \dot{a}\approx -\frac{N}{2 a (2\omega+3)}\left(\frac{\omega \pi_a}{3\phi}+\frac{\pi_\phi}{a}\right)\;,\\
\label{eq_neq32_pi_a}
&&\dot{\pi}_a\approx -\frac{N}{2a^2(2\omega+3)}\left(
\frac{\omega\pi_a^2}{6\phi}+\frac{2\pi_a\pi_\phi}{a}-\frac{3\phi\pi_\phi^2}{a^2}\right)
-3 N a^2 U(\phi)\,,\\
\label{eq_neq32_phi}
&& \dot{\phi}\approx \frac{N}{2 a^2 (2\omega+3)}
\left(-\pi_a+\frac{2\phi \pi_\phi}{a}\right)\,,\\
&&\dot{\pi}_\phi\approx-\frac{N}{2 a (2\omega+3)}\left(\frac{\omega\pi_a^2}{6\phi^2}+\frac{\pi_\phi^2}{a^2}\right)
-Na^3\frac{dU}{d\phi}
\label{eq_neq32_pi_phi}
\end{eqnarray}

\subsection{$\omega=-3/2$ Case}
\label{2.2}

Note that, in this case, the determinant \eqref{Hess} vanishes.
The two definitions for $\pi_a$ and $\pi_\phi$ are not independent; therefore, 
there is an additionally primary constraint:
\begin{equation}
\label{def_C_phi}
C_\phi\equiv \frac{1}{2}a\pi_{a}- \phi \pi_{\phi}\,.  
\end{equation}

The canonical Hamiltonian is:
\begin{eqnarray}
H_{C}^{(-3/2)}\equiv p_i\dot{q}^i-\mathcal{L} =
N\left[ - \frac{\pi_a^2}{24 a \phi}
+ a^3 U(\phi)\right]\,.
\label{hamiltopunto32}
\end{eqnarray}
the Hamiltonian constraint $H^{(-3/2)}$ is the quantity under square brackets.
We use the superscript $\cdots^{(-3/2)}$ in order to emphasize when a 
quantity is evaluated in the  case of $\omega=-\frac{3}{2}$; see also, \cite{Zhang2011,Gielen,galaverni2021jordan}.
We introduce an effective Hamiltonian defined as:
\begin{equation}
 H_{\mathrm{E}}^{(-3/2)}\equiv N H^{(-3/2)} + \lambda_N \pi_N\ + \lambda_\phi C_\phi\,,
\end{equation}
where $\lambda_N(q^i, p_i) $ and $\lambda_\phi$ are Lagrange multipliers.

In order to preserve the primary constraints $\pi_N\approx 0$ and $C_\phi\approx 0$,
we require:
\begin{eqnarray}
\dot{\pi}_N&=&\left\{ \pi_N,  H_{\mathrm{E}}^{(-3/2)} \right\}=-H^{(-3/2)}\approx 0\,,\\
\dot{C}_\phi&=&\left\{ C_\phi,  H_{\mathrm{E}}^{(-3/2)} \right\}
=-N \left[\frac{\pi_a^2}{48 a \phi}-\phi a^3 \frac{d U(\phi)}{d\phi} +\frac{3}{2}a^3 U(\phi)  \right]\nonumber\\
&=& \frac{N}{2}\left[-\frac{\pi_a^2}{24 a \phi}+ a^3 U(\phi)\right]=\frac{1}{2}NH^{(-3/2)}\approx 0\,, 
\end{eqnarray}
where we have used the condition $\phi \frac{d U(\phi)}{d\phi} = 2 U(\phi) $; see Equation \eqref{eqstophi} for $\omega=-\frac{3}{2}$. 
The Hamiltonian constraint $H^{(-3/2)}$ is easily preserved:
\begin{equation}
\dot{H}^{(-3/2)}=\left\{ H^{(-3/2)},  H_{\mathrm{E}}^{(-3/2)} \right\}\approx 0\,.   
\end{equation}
this closes the constraint analysis and there are no further constraints.
Employing all previous calculations, it is quite evident that:
\begin{eqnarray}
    \left\{ \pi_N,  H^{(-3/2)} \right\}=0\, , \left\{ \pi_N,  C_{\phi} \right\}=0\, ,\left\{ C_{\phi},  H^{(-3/2)} \right\}=\frac{1}{2}H^{(-3/2)}\approx 0\,,
\end{eqnarray}
therefore, all constraints are first class. 

The total Hamiltonian is: 
\begin{equation}
H_{\mathrm{T}}^{(-3/2)}\equiv  H_{\mathrm{E}}^{(-3/2)} + \lambda_H H^{(-3/2)}=  N H^{(-3/2)} + \lambda_N \pi_N+ \lambda_\phi C_\phi + \lambda_H H^{(-3/2)} 
\end{equation}
since $\lambda_H$ and $N$ are two arbitrary multipliers, we can write this in a more simple way:
\begin{equation}
H_{\mathrm{T}}^{(-3/2)}=N H^{(-3/2)}  + \lambda_N \pi_N+ \lambda_\phi C_\phi =H_{\mathrm{E}}^{(-3/2)} 
\end{equation}

The equations of motion in the Jordan frame for $\omega=-3/2$ are:
\begin{eqnarray}
\label{eq_eq32_N}
&& \dot{N}\approx \lambda_N \;,\\
\label{eq_eq32_pi_N}
&&\dot{\pi}_N=-H^{(-3/2)}\approx 0 \\
\label{eq_eq32_a}
&& \dot{a}\approx -\frac{N\pi_a}{12 a\phi}+\frac{\lambda_\phi a}{2}\;,\\
\label{eq_eq32_pi_a}
&&\dot{\pi}_a\approx -\frac{N \pi_a^2}{24 \phi a^2}-3 N a^2 U(\phi)-\frac{\lambda_\phi\pi_a}{2}\,,\\
\label{eq_eq32_phi}
&& \dot{\phi}\approx -\lambda_\phi\phi\,,\\
&&\dot{\pi}_\phi\approx-N\frac{\pi_a^2}{24 a\phi^2}-Na^3\frac{dU}{d\phi}+\lambda_\phi \pi_\phi=-N\frac{\pi_a^2}{24 a\phi^2}-\frac{2 N a^3 U}{\phi}+\lambda_\phi \pi_\phi \,.
\label{eq_eq32_pi_phi}
\end{eqnarray}
in the last equation, we have used the condition $\phi \frac{d U(\phi)}{d\phi} = 2 U(\phi) $; see Equation \eqref{eqstophi} for $\omega=-\frac{3}{2}$. 

\section{Anti-Gravity Transformations}
\label{sectIII}
Following \cite{Niedermaier2019,Niedermaier2020,Zhang2011,Zhou2012,Gionti2021,galaverni2021jordan},
we consider the following anti-gravity (or anti-\linebreak Newtonian)~transformations:
\begin{eqnarray}
&&{{\widetilde {N}}^{*}}=N \, \, ; \, {\widetilde{\pi}}^{*}_N=\pi_N  \,\, ; \, \widetilde{a}=(16\pi G\phi)^{1\over 2}a \, \, ; \nonumber \\
&&{\widetilde \pi}_{a}=\frac{{\pi}_{a}}{(16\pi G\phi)^{1\over 2}} \, \,;\,{{\widetilde {\phi}}}=\phi \, \, ; \, {\widetilde \pi}_\phi=\frac{1}{\phi} \, ( \phi \pi_{\phi}-\frac{1}{2}a\pi_{a})\,,
\label{antigravity}
\end{eqnarray}
{These transformations correspond to conformal transformations implemented only on the spatial metric ($g_{ij}\mapsto\lambda^2 g_{ij}$). 
For large values of $\lambda$, these transformations emulate a large Newton constant and also enhance spacelike distances compared to timelike ones \cite{Niedermaier2019}.}

{
One can easily verify that, applying the definition \eqref{canonicalone}, this set of transformations is~canonical:
\begin{eqnarray}
\left\{{\widetilde {N}}^{*},{\widetilde{\pi}}^{*}_N  \right\}
= \left\{\widetilde{a}, \widetilde{\pi}_a\right\}
=\left\{\widetilde{\phi}, \widetilde{\pi}_\phi\right\}=1
\end{eqnarray}
and all of the other Poisson brackets vanish.
}

If we apply these transformations to the Lagrangian \eqref{lagra}, we obtain:
\begin{equation}
\widetilde{\mathcal L}=-\frac{1}{{{\widetilde {N}}^{*}} (16\pi G{\widetilde {\phi}})^{3\over 2}} \left[
{6 \widetilde{a} \dot{\widetilde{a}}^2}{\widetilde {\phi}}-\frac{(2\omega+3)\widetilde{a}^3\dot{{\widetilde {\phi}}}^2}{2{\widetilde {\phi}}}
\right]
-(16\pi G{\widetilde {\phi}})^{1\over 2} {{\widetilde {N}}^{*}} \widetilde{a}^3 V(\widetilde{\phi})
\label{Ltilde}
\end{equation}
where we have introduced: 
\begin{equation}
V(\widetilde{\phi})\equiv\frac{U(\widetilde{\phi})}{(16\pi G\phi)^2}\,.
\label{Vtilde}
\end{equation}

\subsection{$\omega\neq-3/2$ Case}
In the case of $\omega\neq-3/2$,  the canonical Hamiltonian is:
\begin{equation}\begin{array}{ccc}
\widetilde{H}_{C}^*&=&\frac{{\widetilde{{N}}^{*}} {\widetilde{a}}^{3} }{(16\pi G\widetilde{\phi})^{3\over 2}}
\left[ -\frac{\widetilde{\pi}_a^2 2\pi G (16\pi G\widetilde{\phi})^{2} }{3 \widetilde{a}^{4}}
+\frac{\widetilde{\pi}_\phi^2 \widetilde{\phi}^2 8 \pi G (16\pi G\widetilde{\phi})^{2} }{(2\omega +3) \widetilde{a}^{6} } +U(\widetilde{\phi}) \right]\\
&=&\widetilde{N}^* \widetilde{a}^3 (16\pi G\widetilde{\phi})^{1\over 2}  \left[
-\frac{2 \pi G \widetilde{\pi}_a^2}{3\widetilde{a}^4}
+\frac{8\pi G \widetilde{\pi}_\phi^2 \widetilde{\phi}^2 }{(2\omega+3)\widetilde{a}^6} +V(\widetilde{\phi})
\right]
\, .
\label{h_antigravity}
\end{array}
\end{equation}
In a similar way, as discussed before, we recognize that the the total Hamiltonian now is:
\begin{equation}
\widetilde{H}_T^*=  \widetilde{N}^* \widetilde{H}^* + \widetilde{\lambda}^*_N \widetilde{\pi}_N^*  
\end{equation}
where the new Hamiltonian constraint is:
\begin{equation}
\widetilde{H}^*\equiv 
\widetilde{a}^3 (16\pi G\widetilde{\phi})^{1\over 2}  \left[
-\frac{(2 \pi G) \widetilde{\pi}_a^2}{3\widetilde{a}^4}
+\frac{(8\pi G) \widetilde{\pi}_\phi^2 \widetilde{\phi}^2 }{(2\omega+3)\widetilde{a}^6}+V(\widetilde{\phi})
\right]
\end{equation}
Therefore, the equations of motion in the anti-gravity frame for $\omega\neq-3/2$ are:
\begin{eqnarray}
\label{eq_neq32_antig_N}
&& \dot{\widetilde{N}}^*\approx \widetilde{\lambda}_N^* \;,\\
&&\dot{\widetilde{\pi}}_N=-\widetilde{H}^*\approx 0\,, \\
&& \dot{\widetilde{a}}\approx -\widetilde{N}^* (16\pi G \widetilde{\phi})^{1\over 2} \frac{4\pi G \widetilde{\pi}_a }{3 \widetilde{a}}\;,\\
&&\dot{\widetilde{\pi}}_a\approx \widetilde{N}^* (16\pi G \widetilde{\phi})^{1\over 2}
\left[-\frac{(2\pi G) \widetilde{\pi}_a^2 }{3 \widetilde{a}^2}
+\frac{3 (8\pi G)\widetilde{\pi}_\phi^2 \widetilde{\phi}^2 }{(2\omega+3)\widetilde{a}^4}
-3 \widetilde{a}^2 V(\widetilde{\phi}) \right]\,,\\
&& \dot{\widetilde{\phi}}\approx \widetilde{N}^* (16\pi G  \widetilde{\phi})^{1\over 2}
\frac{(16 \pi G \widetilde{\phi})\widetilde{\pi}_\phi \widetilde{\phi}^2 }{(2\omega+3)\widetilde{a}^3}\,,\\
&& 
\dot{\widetilde{\pi}}_\phi\approx \widetilde{N}^* (16\pi G\widetilde{\phi} )^{1\over 2}
\left[
\frac{\pi G \widetilde{\pi}_a^2 }{3 \widetilde{a} \widetilde{\phi}}
- \frac{20 \pi G \widetilde{\pi}_\phi^2 \widetilde{\phi} }{(2\omega+3)\widetilde{a}^3} 
-\frac{\widetilde{a}^3}{2\widetilde{\phi}}V(\widetilde{\phi}) -\widetilde{a}^3 \frac{d V(\widetilde{\phi})}{d\widetilde{\phi}}
\right]
\label{eq_neq32_antig_pi_phi}
\end{eqnarray}
Once we apply the {inverse} anti-gravity transformations, defined in Equation \eqref{antigravity}, we pass from one set of equations of motion, 
Equations \eqref{eq_neq32_antig_N}--\eqref{eq_neq32_antig_pi_phi}, 
to the other,
Equations \eqref{eq_neq32_N}--\eqref{eq_neq32_pi_phi}.
This straightforwardly verifies that the anti-gravity transformations are canonical.

\subsection{$\omega=-3/2$ Case}

In the $\omega=-3/2$ case, if we apply the anti-gravity transformations to the canonical Hamiltonian \eqref{hamiltopunto32}, we obtain:
\begin{eqnarray}
\widetilde{H}_{C}^{*(-3/2)}&=&\frac{{\widetilde{{N}}^{*}} {\widetilde{a}}^{3} }{(16\pi G\widetilde{\phi})^{3\over 2}}
\left[ -\frac{\widetilde{\pi}_a^2 2\pi G (16\pi G\widetilde{\phi})^{2} }{3 \widetilde{a}^{4}}
+U(\widetilde{\phi}) \right]\nonumber\\
&=&\widetilde{N}^* \widetilde{a}^3 (16\pi G\widetilde{\phi})^{1\over 2}  \left[
-\frac{2 \pi G \widetilde{\pi}_a^2}{3\widetilde{a}^4}
 +V(\widetilde{\phi})
\right]
\, .
\label{h_antigravity_32}
\end{eqnarray}
where we have used the definition of $V(\widetilde{\phi})$; see Equation \eqref{Vtilde}.

Looking at the transformed Lagrangian defined in Equation \eqref{Ltilde} for $\omega=-3/2$, it is clear that
the extra primary constraint  is  now ${\widetilde{\pi}}_\phi \approx 0$.
Since we have introduced, according to the common definition in literature \cite{Olmo,Gielen},
$C_\phi\equiv \frac{1}{2}a\pi_{a}- \phi \pi_{\phi}$---see Equation \eqref{def_C_phi}---we have:
\begin{equation}
{\widetilde{\pi}}_\phi=\frac{1}{\phi} \, ( \phi \pi_{\phi}-\frac{1}{2}a\pi_{a})=-\frac{C_\phi}{\phi}\approx0\,
\end{equation}
where we have used the anti-gravity transformations \eqref{antigravity}.
Therefore the primary constraint $C_\phi\approx0$ maps into $\widetilde{C}_\phi\equiv-\widetilde{\phi}\widetilde{\pi}_\phi\approx0.$
The other primary constraint $\pi_N \approx 0$  by the anti-gravity transformations maps into ${\widetilde\pi}_N \approx 0$ 

The total Hamiltonian $\widetilde{H}_{T}^{*(-3/2)}$ is:
\begin{equation}
\widetilde{H}_{T}^{*(-3/2)}=\widetilde{N}^* \widetilde{H}^{*(-3/2)} + \widetilde{\lambda}^*_N \widetilde{\pi}_N^* + \lambda_{\phi} \widetilde{C}_\phi
\end{equation}
The equations of motion in the anti-gravity frame for $\omega=-3/2$ are:
\begin{eqnarray}
\label{eq_eq32_antig_N}
&& \dot{\widetilde{N}}^*\approx \widetilde{\lambda}_N^* \;,\,\dot{\widetilde{\pi}}_N=-\widetilde{H}^{*(-3/2)}\approx 0\,, \\
&& \dot{\widetilde{a}}\approx -\widetilde{N}^* (16\pi G \widetilde{\phi})^{1\over 2} \frac{4\pi G \widetilde{\pi}_a }{3 \widetilde{a}}\;,\\
&&\dot{\widetilde{\pi}}_a\approx \widetilde{N}^* (16\pi G \widetilde{\phi})^{1\over 2}
\left[-\frac{(2\pi G) \widetilde{\pi}_a^2 }{3 \widetilde{a}^2}
-3 \widetilde{a}^2 V(\widetilde{\phi}) \right]\,,\\
&& \dot{\widetilde{\phi}}\approx - \lambda_{\phi} \widetilde{\phi}\,,\\
&& 
\dot{\widetilde{\pi}}_\phi\approx \widetilde{N}^* (16\pi G\widetilde{\phi} )^{1\over 2}
\left[
\frac{\pi G \widetilde{\pi}_a^2 }{3 \widetilde{a} \widetilde{\phi}}
-\frac{\widetilde{a}^3}{2\widetilde{\phi}}V(\widetilde{\phi}) -\widetilde{a}^3 \frac{d V(\widetilde{\phi})}{d\widetilde{\phi}}
\right]+\lambda_{\phi}\widetilde{\pi}_\phi\nonumber\\
&&=\widetilde{N}^* (16\pi G\widetilde{\phi} )^{1\over 2}
\left[
\frac{\pi G \widetilde{\pi}_a^2 }{3 \widetilde{a} \widetilde{\phi}}
-\frac{\widetilde{a}^3}{2\widetilde{\phi}}V(\widetilde{\phi}) 
\right]+\lambda_{\phi}\widetilde{\pi}_\phi
\label{eq_eq32_antig_pi_phi}
\end{eqnarray}
where, in the last equation, we have used the condition $\phi \frac{d U(\phi)}{d\phi} = 2 U(\phi) $ or $\widetilde{\phi}^2 \frac{d V(\widetilde{\phi})}{d\widetilde{\phi}}=0 $; see Equation \eqref{eqstophi} for $\omega=-\frac{3}{2}$. 

Once we apply the inverse anti-gravity transformations, we pass from this set of equations of motion, 
Equations \eqref{eq_eq32_antig_N}--\eqref{eq_eq32_antig_pi_phi}, 
to the set of equations studied in Section \ref{sectII},
Equations \eqref{eq_eq32_N}--\eqref{eq_eq32_pi_phi}.
This straightforwardly verifies that the anti-gravity transformations are also canonical for the case of $\omega=-3/2$.

\section{Transformations from Jordan Frame to the Einstein Frame}
\label{sectIV}

We consider in this section the Weyl (conformal) transformations from the Jordan frame to the Einstein Frame 
\cite{Dyer,Deruelle2009,Gionti2021,galaverni2021jordan}:
\begin{eqnarray}
&&{{\widetilde {N}}}=N (16\pi G\phi)^{1\over 2} \, \, ; \, {\widetilde{\pi}}_N=\frac{\pi_N}{(16\pi G\phi)^{1\over 2}}  \,\, ; \, \widetilde{a}=(16\pi G\phi)^{1\over 2}a \, \, ; \nonumber \\
&&{\widetilde \pi}_{a}=\frac{{\pi}_{a}}{(16\pi G\phi)^{1\over 2}} \, \,;\,{{\widetilde {\phi}}}=\phi \, \, ; \, {\widetilde \pi}_\phi=\frac{1}{\phi} \, ( \phi \pi_{\phi}-\frac{1}{2}a\pi_{a})\,,
\label{JFEFtrans}
\end{eqnarray}
{We easily verify that, applying the definition \eqref{canonicalone}, this set of transformations is not\linebreak \mbox{canonical, since:}
\begin{eqnarray}
\left\{{\widetilde {N}},{\widetilde{\pi}}_\phi  \right\}
=\frac{8\pi G N}{(16\pi G\phi)^{1\over 2}}\neq 0\,.
\end{eqnarray}
}
If we apply these transformations to the Lagrangian \eqref{lagra}, we obtain:
\begin{equation}
\widetilde{\mathcal L}=-\frac{1}{{\widetilde {N}(16\pi G\widetilde{\phi})}} \left[
6 \widetilde{a} \dot{\widetilde{a}}^2 \widetilde{\phi}-\frac{(2\omega+3)\widetilde{a}^3\dot{{\widetilde {\phi}}}^2}{2{\widetilde {\phi}}}
\right]
- {{\widetilde {N}}} \widetilde{a}^3 V(\widetilde{\phi})\,.
\label{Ltilde_JFEF}
\end{equation}

\subsection{$\omega\neq-3/2$ Case}

When we apply Weyl (conformal) transformations for the canonical Hamiltonian, we obtain:
\begin{equation}\begin{array}{ccc}
\widetilde{H}_{C}&=&\frac{{\widetilde{{N}}} {\widetilde{a}}^{3} }{(16\pi G\widetilde{\phi})^{2}}
\left[ -\frac{\widetilde{\pi}_a^2 2\pi G (16\pi G\widetilde{\phi})^{2} }{3 \widetilde{a}^{4}}
+\frac{\widetilde{\pi}_\phi^2 \widetilde{\phi}^2 8 \pi G (16\pi G\widetilde{\phi})^{2} }{(2\omega +3) \widetilde{a}^{6} } +U(\widetilde{\phi}) \right]\\
&=&\widetilde{N} \widetilde{a}^3  \left[
-\frac{2 \pi G \widetilde{\pi}_a^2}{3\widetilde{a}^4}
+\frac{8\pi G \widetilde{\pi}_\phi^2 \widetilde{\phi}^2 }{(2\omega+3)\widetilde{a}^6} +V(\widetilde{\phi})
\right]
\, .
\label{h_JFEF}
\end{array}
\end{equation}
In a similar way, as discussed before, we recognize that the the total Hamiltonian is now:
\begin{equation}
\widetilde{H}_T=  \widetilde{N} \widetilde{H} + \widetilde{\lambda}_N \widetilde{\pi}_N  
\end{equation}
where the new Hamiltonian constraint is:
\begin{equation}
\widetilde{H}\equiv 
\widetilde{a}^3   \left[
-\frac{(2 \pi G) \widetilde{\pi}_a^2}{3\widetilde{a}^4}
+\frac{(8\pi G) \widetilde{\pi}_\phi^2 \widetilde{\phi}^2 }{(2\omega+3)\widetilde{a}^6}+V(\widetilde{\phi})
\right]
\end{equation}
The equations of motion in the Einstein frame for $\omega\neq-3/2$ are:
\begin{eqnarray}
\label{eq_neq32_JFEF_N}
&& \dot{\widetilde{N}}\approx \widetilde{\lambda}_N \;,\\
&&\dot{\widetilde{\pi}}_N=-\widetilde{H}\approx 0\,, \\
&& \dot{\widetilde{a}}\approx -\widetilde{N} \frac{4\pi G \widetilde{\pi}_a }{3 \widetilde{a}}\;,\\
&&\dot{\widetilde{\pi}}_a\approx \widetilde{N} 
\left[-\frac{(2\pi G) \widetilde{\pi}_a^2 }{3 \widetilde{a}^2}
+\frac{3 (8\pi G)\widetilde{\pi}_\phi^2 \widetilde{\phi}^2 }{(2\omega+3)\widetilde{a}^4}
-3 \widetilde{a}^2 V(\widetilde{\phi}) \right]\,,\\
&& \dot{\widetilde{\phi}}\approx \widetilde{N} 
\frac{(16 \pi G )\widetilde{\pi}_\phi \widetilde{\phi}^2 }{(2\omega+3)\widetilde{a}^3}\,,\\
&& 
\dot{\widetilde{\pi}}_\phi\approx -\widetilde{N} 
\left[
 \frac{16 \pi G \widetilde{\pi}_\phi^2 \widetilde{\phi} }{(2\omega+3)\widetilde{a}^3}
+\widetilde{a}^3 \frac{d V(\widetilde{\phi})}{d\widetilde{\phi}}
\right]\,.
\label{eq_neq32_JFEF_pi_phi}
\end{eqnarray}
Once we apply the inverse Weyl (conformal) transformations from the Einstein frame to the Jordan frame, we obtain:
\begin{eqnarray}
\label{eq_neq32_JFEF_inv_N}
&&\color{red}{\dot{N}\approx\frac{\widetilde{\lambda}_N}{(16\pi G\phi)^{\frac{1}{2}}}
-\frac{N^2}{2 a^2(2\omega+3)}\left(\frac{\pi_\phi}{a}-\frac{\pi_a}{2\phi}\right)\;,} \\
\label{eq_neq32_JFEF_inv_pi_N}
&&\color{red}{\dot{\pi}_N\approx-  H+\frac{N\pi_N}{2a^2(2\omega+3)}\left(\frac{\pi_\phi}{a}-\frac{\pi_a}{2\phi}\right)\,,} \\
\label{eq_neq32_JFEF_inv_a}
&& \dot{a}\approx -\frac{N}{2 a (2\omega+3)}\left(\frac{\omega \pi_a}{3\phi}+\frac{\pi_\phi}{a}\right)\;,\\
\label{eq_neq32_JFEF_inv_pi_a}
&&\dot{\pi}_a\approx -\frac{N}{2a^2(2\omega+3)}\left(
\frac{\omega\pi_a^2}{6\phi}+\frac{2\pi_a\pi_\phi}{a}-\frac{3\phi\pi_\phi^2}{a^2}\right)-3 N a^2 U(\phi)\,,\\
\label{eq_neq32_JFEF_inv_phi}
&& \dot{\phi}\approx \frac{N}{2 a^2 (2\omega+3)}\left(-\pi_a+\frac{2\phi \pi_\phi}{a}\right)\,,\\
&&\color{red}{\dot{\pi}_\phi\approx-\frac{N}{2 a (2\omega+3)}\left(\frac{\omega\pi_a^2}{4\phi^2}-\frac{7\pi_\phi^2}{2a^2}+\frac{\pi_a\pi_\phi}{2a\phi}\right)
-Na^3\frac{dU}{d\phi}+\frac{N a^3}{2\phi}U(\phi)\,.}
\label{eq_neq32_JFEF_inv_pi_phi}
\end{eqnarray}
Note that Equations \eqref{eq_neq32_JFEF_inv_N}, \eqref{eq_neq32_JFEF_inv_pi_N}, and \eqref{eq_neq32_JFEF_inv_pi_phi} (in red colour) do not coincide with the corresponding equations in the Jordan frame, as in Section \ref{2.1}. This is a consequence of the Hamiltonian non-canonicity of the transformations from the Jordan to Einstein frames

\subsection{$\omega=-3/2$ Case}
When we apply Weyl (conformal) transformations defined in Equation \eqref{JFEFtrans} for the canonical Hamiltonian, we obtain:
\begin{eqnarray}
\widetilde{H}_{C}^{(-3/2)}&=&\frac{{\widetilde{{N}}} {\widetilde{a}}^{3} }{(16\pi G\widetilde{\phi})^{2}}
\left[ -\frac{\widetilde{\pi}_a^2 2\pi G (16\pi G\widetilde{\phi})^{2} }{3 \widetilde{a}^{4}}
 +U(\widetilde{\phi}) \right]\nonumber\\
&=&\widetilde{N} \widetilde{a}^3  \left[
-\frac{2 \pi G \widetilde{\pi}_a^2}{3\widetilde{a}^4}
+V(\widetilde{\phi})
\right]
\, .
\end{eqnarray}
In a similar way, as discussed in the previous section, we recognize that the the total Hamiltonian now is:
\begin{equation}
\widetilde{H}_T^{(-3/2)}=  \widetilde{N} \widetilde{H}^{(-3/2)} + \widetilde{\lambda}_N \widetilde{\pi}_N +\lambda_\phi\widetilde{C}_\phi\,, 
\end{equation}
where the new Hamiltonian constraint is:
\begin{equation}
\widetilde{H}^{(-3/2)}\equiv 
\widetilde{a}^3   \left[
-\frac{(2 \pi G) \widetilde{\pi}_a^2}{3\widetilde{a}^4}
+V(\widetilde{\phi})
\right]\,.
\end{equation}
The equations of motion in the Einstein frame for $\omega=-3/2$ are:
\begin{eqnarray}
&& \dot{\widetilde{N}}\approx \widetilde{\lambda}_N \;,\\
&&\dot{\widetilde{\pi}}_N=-\widetilde{H}\approx 0\,, \\
&& \dot{\widetilde{a}}\approx -\widetilde{N} \frac{4\pi G \widetilde{\pi}_a }{3 \widetilde{a}}\;,\\
&&\dot{\widetilde{\pi}}_a\approx \widetilde{N} 
\left[-\frac{(2\pi G) \widetilde{\pi}_a^2 }{3 \widetilde{a}^2}
-3 \widetilde{a}^2 V(\widetilde{\phi}) \right]\,,\\
&& \dot{\widetilde{\phi}}\approx -\lambda_\phi\widetilde{\phi}\,,\\
&& 
\dot{\widetilde{\pi}}_\phi\approx -\widetilde{N} \widetilde{a}^3 \frac{d V(\widetilde{\phi})}{d\widetilde{\phi}}
+\lambda_\phi \widetilde{\pi}_\phi=\lambda_\phi \widetilde{\pi}_\phi\,.
\end{eqnarray}
where, in the last equation, we have used the condition $\phi \frac{d U(\phi)}{d\phi} = 2 U(\phi) $ or $\widetilde{\phi}^2 \frac{d V(\phi)}{d\widetilde{\phi}}=0 $; see Eq. for $\omega=-\frac{3}{2}$.

Once we apply the inverse Weyl (conformal) transformations from the Jordan frame to the Einstein frame, 
we obtain:
\begin{eqnarray}
\label{eq_eq32_JFEF_inv_N}
&&\color{red}{\dot{N}\approx \frac{\widetilde{\lambda}_N}{(16\pi G\phi)^{1\over 2}}+\frac{\lambda_\phi N}{2} } \;,\\
\label{eq_eq32_JFEF_inv_pi_N}
&&\color{red}{\dot{\pi}_N\approx-\left(-\frac{\pi_a^2}{24 a \phi}+a^3 U(\phi)\right)-\frac{\lambda_\phi\pi_N}{2}=-H^{(-3/2)}-\frac{\lambda_\phi\pi_N}{2}}\,, \\
\label{eq_eq32_JFEF_inv_a}
&& \dot{a}\approx -\frac{N\pi_a}{12 a\phi}+\frac{\lambda_\phi a}{2}\;,\\
\label{eq_eq32_JFEF_inv_pi_a}
&&\dot{\pi}_a\approx -\frac{N \pi_a^2}{24 \phi a^2}-3 N a^2 U(\phi)-\frac{\lambda_\phi\pi_a}{2}\,,\\
\label{eq_eq32_JFEF_inv_phi}
&& \dot{\phi}\approx -\lambda_\phi\phi\,,\\
\label{eq_eq32_JFEF_inv_pi_phi}
&&
\color{red}{\dot{\pi}_\phi\approx-N\frac{\pi_a^2}{16 a\phi^2}-\frac{3 N a^3 U}{2\phi}+\lambda_\phi \pi_\phi
=\frac{3}{4}\left(-N\frac{\pi_a^2}{12 a\phi^2}-\frac{2 N a^3 U}{\phi}\right)+\lambda_\phi \pi_\phi \,.}
\end{eqnarray}
Again, we denote in red colour the equations of motion \eqref{eq_eq32_JFEF_inv_N}, \eqref{eq_eq32_JFEF_inv_pi_N}, and \eqref{eq_eq32_JFEF_inv_pi_phi},
not coinciding with the corresponding equations in Section \ref{2.2} as a consequence of the Hamiltonian non-canonicity of the transformations from the Jordan to Einstein frame.

\section{Conclusions}
\label{sectV}

We have carried out  a detailed analysis of a flat mini-superspace model of the FLRW Universe in the Brans--Dicke theory for both cases of $\omega\neq -\frac{3}{2}$ and $\omega= -\frac{3}{2}$. 
The entire analysis highlights, quite clearly, that the Hamiltonian equations of motion in the Jordan frame and in the Einstein frame are not equivalent. In particular, the Hamiltonian  transformations from the Jordan to the Einstein frames  are not  Hamiltonian canonical transformations. Of course, this in-equivalence, at the Hamiltonian level, poses the same question at the Lagrangian level: are the equations of motion in the Jordan frame equivalent to the analogous ones in the Einstein frame? The answer, driven by all considerations that we have made, would be no. Therefore, a further detailed analysis of the Lagrangian equations of motion of the FLRW mini-superspace model in the Brans--Dicke theory appears to be a plausible future project. Tightly connected to it, there is the question of whether Jordan and Einstein frames are really physically equivalent. 
In fact, in the literature, there exist articles claiming their physical equivalence \cite{Deruelle2010,Francfort2019,Frion2018}, as well as their in-equivalence \cite{Bombacigno:2019nua,Capozziello2010,Carloni:2010rfq}. 

\acknowledgments{We thank Alfio Bonanno, Sante Carloni and Ugo Moschella for useful discussions.}

\bibliography{bransdickepartcase}

\end{document}